\documentclass[12pt, preprint]{aastex}

\newcommand{\lunits}{$\rm erg~s^{-1}$}
\newcommand{\funits}{$\rm erg~cm^{-2}~s^{-1}$}
\newcommand{\cunits}{$\rm cm^{-2}$}
\newcommand{\chandra}{{\it Chandra~}}
\newcommand{\rosat}{{\it ROSAT~}}
\newcommand{\xmm}{{\it XMM-Newton~}}
\newcommand{\iras}{IRAS20051-1117}

\makeatletter

\newenvironment{inlinefigure}{%
\def\@captype{inlinefigure}%
\noindent\begin{minipage}{\linewidth}\begin{center}}
{\end{center}\end{minipage}\smallskip}
\makeatother

\shorttitle{X-ray Observations of IRAS20051-1117}
\shortauthors{Georgantopoulos et al.}

\begin{document}
\title{X-ray luminous galaxies II.
 Chandra and XMM-Newton Observations of the 'composite' 
 galaxy IRAS20051-1117}

\author{I.~Georgantopoulos$\!$\altaffilmark{1}
I. Papadakis$\!$\altaffilmark{2}  
A.~Zezas$\!$\altaffilmark{3} 
M.~J.~Ward$\!$\altaffilmark{4} 
}

\altaffiltext{1}{Institute of Astronomy \& Astrophysics, 
National Observatory of Athens, Palaia Penteli, 15236, Athens, Greece}
\altaffiltext{2}{Physics Department, University of Crete,  
71003, Heraklion, Greece}
\altaffiltext{3}{Harvard-Smithsonian Center for Astrophysics, 60 
 Garden Street, Cambridge, MA02138}
\altaffiltext{4}{X-ray Astronomy Group, Department of Physics \&
 Astronomy, University of Leicester, LE1 7RH, UK}

\begin{abstract}
We present {\it Chandra} and {\it XMM-Newton} observations of the
composite star-forming/Seyfert galaxy IRAS20051-1117. 
The X-ray imaging and spectral properties
reveal the presence of an active nucleus. The {\it Chandra} image shows a
strong nuclear point source ($\rm L(2-10 keV) \sim 4\times10^{42} \rm
erg~s^{-1}$). The nuclear X-ray source coincides with a bright, also
un-resolved, UV source which appears in the \xmm\ Optical Monitor images.  
The \xmm\ and \chandra spectrum is well represented by a power-law 
 with a photon index of
$\Gamma\sim 1.7-1.9$ and a thermal component with 
 a temperature of $\sim 0.2$ keV. We also detect an Fe line at 6.4 keV with an equivalent
width of $\sim 0.3$ keV, typical of the iron lines
that have been detected in the X-ray spectra of ``classical" AGN. We find
no evidence for short-term variability in the X-ray light curves, while we
detect no variations between the \xmm and \chandra observations which are
separated by about 20 days.  Optical spectroscopic observations which were
performed $\sim 2.5$ months after the \xmm\ observation show that the
optical spectrum is dominated by a star-forming galaxy component, although
a weak broad $H\alpha$ component is present, in agreement with the results
from past observations. The lack of strong AGN signatures in the optical
spectrum of the source can be explained by the dilution of the nuclear AGN
emission by the nuclear star-forming component and the strong emission of
the underlying, bright host galaxy.
\end{abstract}


\section{Introduction}
 \chandra and \xmm surveys revealed the presence 
 of a large population of 'optically inactive'  or 'passive' 
 galaxies (Mushotzky et al. 2000, 
 Severgnini et al. 2003, Alexander et al. 2003, Green et al. 2004), 
 confirming previous findings by \rosat (Griffiths et al. 1995,
 Boyle et al. 1995, McHardy et al. 1998).
 The optical spectra of these galaxies do not show evidence 
 for either broad or high excitation lines (e.g. Barger et al. 2001). 
 However, in most cases the X-ray luminosities 
 observed are in excess of $\rm L_x \sim 10^{42}erg~s^{-1}$ 
 which is considered as the upper limit for the X-ray emission 
 of bona-fide star-forming galaxies (e.g. Zezas, Georgantopoulos 
 \& Ward 1998, Moran et al. 1999) suggesting the  presence of active nuclei. 
 It is possible that these galaxies represent a mixed class of objects.
 In the case of 'Fiore P3', which 
 is probably the most well studied case of the
 'passive' galaxies,  Comastri et al. (2002) find that it 
 is most likely associated with an obscured AGN. 
 Severgnini et al. (2003) find instead that
 a number of  galaxies detected in their \xmm survey 
 present X-ray spectra with no evidence for absorption. 
 The dichotomy between the optical and X-ray 
 classification could then be explained if the 
 galaxy light dilutes the weak nuclear component
 (see also  Moran et al. 2002, Georgantopoulos et al. 
 2003 for further examples of such objects).    

 The 'optically inactive' galaxies bear many 
 similarities to the 'composite' 
 class of objects discovered by Veron et al. (1981)
 and further discussed by Moran et al. (1996). 
 The diagnostic emission line diagrams
 of Veilleux et al. (1995) classify
 these as star-forming galaxies. Yet, their X-ray luminosities 
 usually exceed $10^{42}$ $\rm erg~s^{-1}$ (Moran et al. 1996);
 moreover, a fraction of the 'composites' present weak $H_\alpha$ broad wings. 
 \chandra observations of the 'composite' IRAS00317-2142 (Georgantopoulos 
   et al. 2003) showed that the X-ray emission of this object is
 dominated  by a variable nuclear point source 
 which presents an unabsorbed X-ray spectrum.
  Here, we present \xmm and \chandra observations 
 of \iras.   This is one of the most luminous 'composite' objects 
 in the Moran et al. (1996) sample, 
 with a high, soft X-ray luminosity 
 (in excess of $10^{42}$ \lunits) 
 at a redshift of $z \approx 0.0315$. 
  The diagnostic emission line ratios
 ($H_\alpha/[NII]$ versus $H_\beta/[OIII]$)  
 classify it as a star-forming galaxy 
 (Moran et al. 1996). Nevertheless, its high 
 X-ray luminosity  as well as a faint broad $H\alpha$ wing 
 would classify this as an AGN.
 The excellent spatial resolution of
 {\it Chandra} combined with the high effective area 
 and good energy resolution of the \xmm mission 
 can be used  to unveil the nature of the AGN which is
 hosted in this galaxy.  
 Throughout this paper, we adopt  
 a Hubble constant of $\rm H_o= 65 km~s^{-1}~Mpc^{-1}$.

\section{Observations and Data Reduction}
 \subsection{Chandra}
 \iras \ was observed using 
 the Advanced CCD Imaging Spectrometer, ACIS-S,
  onboard {\it Chandra} 
 (Weisskopf et al. 1996). The observation date 
 was 20 April 2002 (Sequence Number 700383). 
 We use the cleaned event file provided 
 by the standard pipeline processing  
 having an exposure time of 20.1 ksec. 
  Only Grade 0,2,3,4 and 6 events are used in the analysis.     
 The whole galaxy falls on only one chip (S3).
 Charge Transfer Inefficiency (CTI) problems do not 
 affect our observations as S3 is back-illuminated chip.  
 Each CCD chip subtends an 8.3 arcmin square on the sky
 while the pixel size is $0.5''$. 
 The spatial resolution on-axis is $0.5''$ FWHM. 
 The ACIS-S spectral resolution is 
 $\sim$100 eV (FWHM) at 1.5 keV.
 We observed in  $1/4$-subarray mode  
 in order to minimize pile-up problems; 
 in this mode the size of the chip is reduced 
 effectively to about $\rm 2x8$  arcmin.
 We obtain about 7000 counts in the total 
 0.3-8 keV band, after background subtraction, corresponding to a count rate 
 of 0.35 $\rm cts~s^{-1}$. Then, the {\sl PIMMS} software 
 estimates  a pile-up fraction of less than 10\%.    
 Images, spectra, ancillary files, response matrices, 
 and  lightcurves have been created  
 using the {\sl CIAO v2.2} software.
 We use a $2''$ radius extraction region in order to produce 
 both the spectral files and the lightcurves.   
 We take into account the degradation of the ACIS quantum 
 efficiency in low energies, due to molecular contamination, 
 by using the {\sl ACISABS} model in the 
 spectral fitting\footnote{http://asc.harvard.edu/cal/Acis/Cal\_prods/qeDeg}.
 The imaging analysis was performed using 
 the {\sl SHERPA} software. 
    
\subsection{XMM-Newton} 
 The \xmm data 
 were obtained in 01 April 2002   as part of the
 Guest Observer Programme AO-1.  The EPIC (European Photon Imaging Camera;
 Str\"uder et al. 2001 and Turner et al. 2001) cameras were operated
 in full frame mode with the thin filter applied.
 The data have been cleaned and  processed in the standard 
 pipeline mode using the  Science Analysis
 Software ({\sc sas}). The event files for the PN and the two MOS
 detectors were produced using the standard reduction pipeline
 (Watson et al. 2001). The event files were screened for high
 particle background periods resulting in a good time interval of
 about 5.5 ksec for both MOS and PN. 
 Events corresponding to patterns 0--4 and 0--12
 have been included in the analysis of the PN and MOS data
 respectively. We obtain about 5.500 and 3500 background subtracted 
 counts in the PN and the combined MOS respectively.
 The resulting count rates are $\approx$ 1 and 
 0.7 $\rm cts~s^{-1}$ respectively. 
 We use a 4 pixel 
 ($\sim 16''$) radius extraction region, in order to obtain 
 the spectrum (and the lightcurve) of our source.  
 We obtain the background from a
 region from the same CCD chip with an area about 10 times larger than the
source extraction region.  We create the 
 auxiliary files for the PN and MOS using the {\sc SAS} task {\sc arfgen} 
 to take into account
 both instrumental effects (e.g. quantum efficiency, telescope
 effective area, filter transmission)  and the fraction of light
 falling outside the extraction radius. The spectral
 response files are created using the task {\sc rmfgen}.  

 Finally, during the \xmm\ observation, the OM (Optical Monitor) 
 was operated in Imaging mode.  
The source was observed with the B, UVW1 (i.e. U) and UVW2 filters. The
total exposure time was 4, 5 and 6 ksec respectively. In the subsequent
analysis we use the processed, standard pipeline mode data (which were
reduced using {\sc SAS} version 5.3.2).

\section{Data Analysis}

\subsection{Imaging}
 In Fig. 1  we plot the \chandra X-ray contours in the total band 
 (0.3-8 keV) overlayed on the \xmm OM (Optical Monitor) image
 in the U (UVW1, 2900 \AA) filter.  All the X-ray emission emanates from the 
 source coinciding with the optical nucleus. 
 The coordinates of the X-ray source are 
  $\alpha = 20h07m51.3s$, $\delta =-11^\circ 08m33s$ (J2000); note that 
 the {\it IRAS}  name  for this object refers to B1950.   

 We have fitted the two-dimensional spatial profile of the
 \chandra source with the {\sc SHERPA} software.
 Fitting a Gaussian profile, we obtain a FWHM of $1.78\pm 0.02$ pixels.
 Given that the nominal Point Spread Function (PSF) is undersampled by the
 pixels of the ACIS camera (FWHM$_{PSF}\sim0.5''$ at 1.5~keV,
 compared to a pixel size of  $0.49''$)
 we consider the probability of any extension marginal.
 Indeed, the Gaussian fit to the Point Spread Function 
 generated with the {\sc CIAO mkpsf} task  gives a comparable 
 FWHM ($\approx 1.5\pm 0.02$ pixels). 
 
Fig. 1 shows that the central UV-bright source in \iras\ 
is surrounded by some faint nebulosity which is associated with 
the galaxy. The nuclear source appears to be point-like in the UV. 
  In order to investigate this issue further, we
co-added the 5 UVW1 and UVW2 images to create a single image. Using {\sc
IRAF}, we fitted the one-dimensional radial profile of the central source
in both filters with a Gaussian. The FWHM of the Gaussian profile is $\sim
2.5''$ and $\sim 1.8''$ in the case of the UVW1 and UVW2 images,
respectively. The profile of the UVW1 source appears to be slightly larger
than the width of the nominal PSF for this filter ($1.7''$). The FWHM of
the UVW2 source is identical to the PSF width for this filter ($1.8''$).
As the contribution of the underlying galaxy is much more pronounced in
the UVW1 image, our results imply that, just like with the X-ray source,
the UV source is not resolved either, its size being equal to or smaller
than $\sim 2''$ which corresponds to $\sim$1.4 kpc.

\subsection{Timing}

Fig. 2 (upper panel) shows the background subtracted PN and MOS light curves for
\iras\ in the 0.3 - 8 keV band (open squares, filled and open circles for
PN, MOS-1 and MOS-2 light curves, respectively). The time bin size was set
to 200 sec. All cameras were switched on about the same time, however
there are a few gaps of duration $\sim 20-800$ sec in all light curves,
 due to the rejetion of periods with high particle background. 
 The light curves appear to be flat, with no
significant variations above the intrumental noise. Using the $\chi^{2}$
test, we find that the light curves are consistent with the hypothesis of a
constant flux. This is true even when we consider the combined, background
subtracted PN and MOS light curves. The \chandra\ 0.3-8 keV light curve
does not show any significant, intrinsic variations either (Fig. 2, lower panel). 
We conclude that \iras\ does not show any significant variations on time scales
smaller than $\sim 10$ ksec.

 Next, we examine whether there is any evidence for long term variability.
 Using the best fit power-law spectrum derived from the PN data 
($\Gamma=1.94$ see section 3.3 below) we derive a 2-10 keV flux 
 of $1.5\pm0.025\times10^{-12}$ and $1.5\pm0.020\times10^{-12}$ 
 \funits from the \xmm (PN) and \chandra data respectively, 
 suggesting that the nucleus presents no variability.    
 However, as the \chandra and \xmm observations are separated by a small 
 interval of about 3 weeks, we have also checked whether 
 other X-ray missions have obtained in the past observations of our object.
 {\it ROSAT} (PSPC) has observed \iras \ in the second semester 
 of 1990 during the ROSAT-All-Sky-Survey (exposure time 411s). 
 The count rate of $0.07\pm 0.01$ (0.1-2.4 keV) translates to 
  an absorbed flux of $8.4\pm1.2 \times10^{-13}$ \funits in the 0.3-2 keV band
 (using the best-fit PN model). This is consistent  
  with the observed \xmm and \chandra flux in 
 the same energy band ($\sim 1\times 10^{-12}$ \funits),
 within $\rm \sim 1.3\sigma$.
 
\subsection{The X-ray spectra} 
  
The spectra are grouped, using the {\sc ftool} {\sc grppha} task to give a
minimum of 20 (15) counts per bin for \xmm (\chandra)  to ensure that
Gaussian statistics apply. The quoted errors to the best fitting spectral
parameters correspond to the 90 \% confidence level for one interesting
parameter. In order to assess the significance of new parameters
added to a model we have adopted the 95 \%  confidence level, using
the {\it F}-test for additional terms. We discard data below 0.3 and
above 8 keV due to the low response of the ACIS-S, PN and MOS cameras.  
We fit the data using the {\sc XSPEC v11.2} software package. The
spectral fits are presented separately for the \chandra and the \xmm data.

\underline{a) The \xmm spectral fits.} We first fit the PN data using a
single power-law model absorbed by two columns: one fixed to the Galactic
($\rm N_H \approx 7\times10^{20}$ Dickey \& Lockman 1990)  and one
intrinsic which is left free to vary.  We find that this model provides a
relatively poor fit ($\chi^2=268.7/190$)  to the data.  The power-law has
the 'canonical' slope, $\Gamma= 1.92^{+0.08}_{-0.07}$, while the 90 \%
upper limit of the intrinsic column is $<1\times10^{20}\rm cm^{-2}$.

The addition of a narrow line ($\sigma<0.01$ keV i.e. smaller than the
CCD's spectral resolution at these energies)  at 6.4 keV (rest-frame)  
results in a better fit. We find $\Delta \chi^2=10.3$ for one additional
parameter (the normalization of the Gaussian line) which is statistically
significant at the $>99$ \% confidence level. The equivalent width of the
 Fe line is $360^{+174}_{-170}$ eV consistent with 
 that predicted for unobscured AGN (e.g. Page et al. 2004)
 with comparable X-ray luminosity. 
 The spectral results are presented in
table 1.  In Fig. 3 (upper panel) we plot the model fit to the data together with
the data to model ratio.  Inspection of the residuals suggests the possibility that
some additional lines are present: e.g. Fe at 6.96 keV, Ca at 3.9 keV, Si
at 1.9 keV (rest-frame).  However, only the first line (Fe)  is detected
at over the 95 per cent confidence level. The resulting equivalent width
is $371^{+263}_{-206}$ eV. The presence of residuals at low 
 energies prompted us to test for the presence of thermal 
 emission. The inclusion of a Raymond-Smith 
 spectrum (abundance fixed at Z=0.3) results in a  significantly better fit 
 $\chi^2 \approx 233.3/186$. The temperature of this component  
 ($\rm kT=0.20^{+0.06}_{-0.02}$ keV) 
 is consistent with the properties of hot gas emission 
 in nearby star-forming galaxies (e.g. Stevens et al. 2003).
 Its luminosity is $\rm L_{0.3-8 keV} \sim 2\times 10^{41}$ \lunits.  
      
The MOS data are fitted with the normalizations untied between the MOS1
and MOS2 detectors. The single power-law model gives a good fit
($\chi^2=128.5/121$) with a photon index $\Gamma=2.0_{-0.09}^{+0.11}$ and
an intrinsic column density of $5^{+2}_{-1}\times10^{20}$ \cunits.  The low
effective area of the MOS detectors at high energies, hampers us from 
 placing any constraints on the properties of the Fe lines above 6 keV. 
The addition of an Fe line results in the reduction of the $\chi^2$ by only $0.1$.
 Moreover, the inclusion of a Raymond-Smith component is not 
 statistically significant. 
Finally, a joint fit to the PN and MOS data (with the  
 normalizations between the 3 detectors untied) yields
 $\Gamma=1.96^{+0.06}_{-0.06}$, $\rm N_H=2^{+1}_{-1}\times10^{20}$ \cunits
 ($\chi^2=381.9/313$). The normalizations between the 2 MOS instruments agree 
 within 5\% while those between MOS and PN within 20 \%.
 The inclusion of the two Fe lines and the Raymond-Smith component 
 is statistically significant yielding 
  $\Gamma=1.96^{+0.08}_{-0.06}$, $\rm N_H=3^{+2}_{-1}\times10^{20}$ \cunits
 ($\chi^2=347.9/307$).

\underline{b) The \chandra spectral fits.} 
 A single power-law fit to the data gives $\Gamma=1.79\pm 0.04$ with 
 a negligible intrinsic column density of $0.1\times 10^{20}$ \cunits
 ($\chi^2_{\nu} =290.5/202$). 
 The above values are in reasonable agreement with those derived 
 from the PN.  The addition
 of an Fe line at 6.4 keV reduces the $\chi^2$ by 6.9 which is statistically
 significant at the $\approx 97$ \% confidence level. 
 The equivalent width of the line is $250\pm 155$ eV, 
 comparable to the value found from the model fitting of the PN data.
 The \chandra spectrum is plotted in Fig. 3 (lower panel) together 
 with the data to model ratio. 
 Finally, the addition of a Raymond-Smith component 
 results in the improvement of the fit ($\chi^2\approx 246.3/199$). 
 The temperature of the Raymond-Smith component is 
 $0.22\pm 0.03$ keV in good agreement 
 with \xmm while the power-law photon 
 index is  flatter ($1.67\pm0.05$). 
 The reduction of the $\chi^2$, when the thermal 
 component is added, may suggest the presence of a compact 
 nuclear star-forming region. Alternatively, 
 it is possible that many {\it photo-ionization} 
 line residuals at soft energies  
 are reduced with the inclusion of the 
 above component.

\subsection{Optical Observations}
                                                                                
Optical spectroscopic observations were made with the 1.3 m telescope at
Skinakas Observatory (Crete, Greece) on 2002 June 21. The f/7.7
Ritchey-Cretien telescope was equipped with a $1024\times1024$ SITe chip
plus a 651 lines mm$^{-1}$ grating, giving a dispersion of $\sim 3.3$~\AA/pix.
We used a 2$''$ slit which corresponds to a physical size of $1.4$ kpc at
the redshift of our object. The spectra span the $4300-7400$ \AA\ range,
at a resolution of $\sim$ 7.0 \AA FWHM. The total integration time was 1.5 hrs.

The optical spectrum of the source is shown in Fig. 4. It is
similar to the spectrum presented by Moran et al. (1996). We detect strong
H$\alpha$, [N II], [S II] and [O III] and much weaker H$\beta$ (narrow) 
and [O I] lines. H$\alpha$ has a broad component with a width of $\sim 3800$
km/sec. Our measurements of the galaxy emission line flux ratios are as
follows: [O III]/H$\beta$(narrow)=1.1, [O I]/H$\alpha$(narrow)=0.019,
H$\alpha$(narrow)/H$\beta$(narrow)=2.65, H$\alpha$(broad)/H$\beta$(narrow) =3.69, 
[NII]/H$\alpha$(narrow)=0.59, and [SII]/H$\alpha$(narrow)=0.33. The ratios
for the bright lines are slightly smaller (by a factor of $\sim 1.5-2.5$)
than the respective values listed in Table 3 of Moran et al. (1996). 
These differences are not highly significant, specially if one considers 
 the fact that different ratio 
values can be obtained if the seeing and/or extraction apertures differ 
since in this case different amounts of the host galaxy are included in 
the aperture. Furthermore, for the
fainter lines (H$\beta$ and [O I]) our measurements are quite uncertain
because of the poor quality of the data (the overall signal-to-noise of
the optical spectrum is $\sim 5$).  Despite these small differences, the
position of the source on the  standard emission line diagnostic diagrams
(e.g. Veilleux et al. 1995)   confirms that this object lies between the
HII region and AGN loci.

From the ratio between H$\alpha$(narrow) and H$\beta$, we calculate a
reddening of $\rm E(B-V)=-0.08$ and $\rm E(B-V)=0.002$ 
 for an AGN and an HII region
intrinsic continuum respectively. These correspond to an equivalent HI
column density of $4\times10^{20}~\rm{cm^{-2}}$ and
$1\times10^{19}~\rm{cm^{-2}}$, assuming Galactic gas-to-dust ratio
(Bohlin et al 1978). As the reddening mentioned above refers to the sum of
the Galactic and intrinsic extinction, this result agrees well with the
result of no intrinsic absorption as derived from the model fitting to the
EPIC PN X-ray spectrum.

\subsection{The spectral energy distribution (SED)}

The availability of simultaneous UV (OM) and X-ray (EPIC-PN)  
observations by \xmm\ gives us the opportunity to investigate the UV/X-ray
energy distribution of the source. For the OM data, we first estimated the
mean count rate of the individual images (as formed by summing the counts
within a $16''$ diameter aperture centered on the nucleus). We have not
corrected for the contribution of the underlying galaxy, which increases
towards larger wavelengths (reaching $\sim 30$ \% in the B filter).
We transformed the average count rates to magnitudes using the appropriate
'Zeropoint magnitude' value for each filter (as listed in the \xmm\ users
manual). For the reddening correction we assumed no intrinsic absorption,
and adopted the Galactic extinction value, $A_{V}\approx0.4$, of Schlegel et
al. (1998).  Using the $A_{\lambda}$ versus $\lambda$ relationship of
Cardelli et al. (1989) for $R_{V}=3.1$, we then found $A_{\lambda}$ at
$\lambda=4400,$ 2900, and 2150 \AA\ (i.e. the wavelengths at which the
response of the B, UVW1, and UVW2 filters peaks).

The UV/X-ray energy distribution is shown in Fig. 5, plotted as
$\nu f_{\nu}$ versus $\nu$.  For the X-ray data we used the best fitting
power law model to the \xmm\ EPIC-PN spectrum (the long dashed lines
represent power law models with slopes equal to the best fitting slope
$\pm 1\sigma$ confidence limit values). 
In the UV band, the ratio $\nu f_{UVW2}/\nu f_{UVW1}$ is 
comparable with the ratio of the respective frequencies. This result 
implies that $\nu f_{nu}\propto \nu$, 
 in agreement with the trend shown by the mean energy
distribution of radio-quiet AGN in the same energy band (e.g. Elvis et al.
1994). Using the data shown in Fig. 5, we find that the
monochromatic X-ray (2 keV) to UV (2500 \AA)  luminosity ratio for \iras\
is $\alpha_{\rm ox}=1.42$. This is consistent with the average value of
$\sim 1.6$ for radio-quiet AGN (Stocke et al. 1991).

The OM, $B$ band measurement indicates that the SED rises towards
longer wavelengths. This trend is caused by the fact the host galaxy's
underlying emission in this band is quite stronger than the emission in the UV
wavelengths. A quasi-simultaneous flux measurement at longer wavelengths
is also possible based on the optical spectrum of the source. Since this
spectrum was taken only $\sim 2.5$ months after the
\xmm\ observation we used it in order to estimate a ``V"  band
measurement. To that aim, we computed the average flux in the $5400-5600$
\AA\ band (observer's frame), which is free of lines, and we dereddened it
adopting $A_{V}\approx 0.4$, as before. 
Interestingly, although the optical spectrum was taken using a $2''$
wide slit (i.e. significantly smaller than the aperture size that we use
for the OM photometry), the $V$ band flux that we measure is larger than
that in the $B$ band. This result gives further support to the idea that
the SED rises towards longer wavelengths.

This trend is confirmed when we consider the near and far infrared
measurements of the source. For the J, H, and K bands we used the 2MASS
measurements (in a aperture of size $14''\times 14''$). For the
far-infrared band we used the {\sc IRAS} measurements at $25, 60,$ and 100
microns, from the {\sc IRAS} faint source catalogue. 
Although the near-infrared and UV flux
measurements are taken from apertures which have almost identical areas,
the strong rise from optical to near-infrared wavelengths shown in
Fig.~5 shows that the energy release in the near-infrared band is
significantly larger than that released in the UV band.
In \iras, it is the near-infrared/optical which rises above
its UV continuum, and not the other way around, as it is observed
in most galaxies with a strong active nucleus.

Apart from the strong contribution of the galaxy stellar population in the
near-infrared bands, the strong far infrared emission is indicative of the
presence of a significant star formation rate (SFR). For example, using
the relation between logarithm of SFR and logarithm of galaxy luminosity
at 100 microns of Misiriotis et al. (2004), we find an overall SFR of
$\sim 32$ M$_{\odot}$ yr$^{-1}$.  This is typical of bright {\sc IRAS}
galaxies, but much smaller than SFR in starburst galaxies.

\section{Discussion}

In this work, we present the results from X-ray and optical observations
of \iras. The \chandra\ and \xmm\ OM images reveal the presense of a
strong, UV and X-ray bright nuclear source, which is un-resolved in both
cases. The FWHM of the X-ray source is $\sim 1''$, which corresponds to a
region of size $\sim 0.7$ kpc at the distance of the object. The X-ray
spectrum of the nucleus is very similar to the typical AGN spectra: a
power-law model fitting to the EPIC-PN spectrum shows the canonical
spectral slope of $\Gamma \sim 1.9$, and reveals the presence of an iron
line at 6.4 keV.  The equivalent width of the line is consistent with 
 that of unobscured AGN of comparable luminosity (e.g. Page et al. 2004).
We also detect evidence for the presence of ionized material in the
vicinity of the nucleus, as indicated by the presence in the spectrum of a
line at 6.96 keV which we interpret as emission from ionized iron.
 Thermal emission with a low temperature (0.2 keV) is present
 in both the \xmm and the \chandra spectrum.  

The presence of a strong, point-like nuclear UV and X-ray source, with an
X-ray spectrum typical of AGN, shows conclusively that the intense X-ray
emission of \iras\ is produced by a Seyfert-like nucleus in the center of
this galaxy. This is further supported by the fact that the UV/X-ray
energy spectral distribution of this object is very similar to the SEDs of
AGN in the same energy band. In particular, the ``X-ray loudness" of this
object, $\alpha_{\rm ox}\approx 1.4$, is consistent with the average
X-ray/UV luminosity ratio found for radio-quiet AGNs.

However, optical observations that were performed only 2.5 months after
the \xmm\ observations of \iras, reveal an optical spectrum very similar
to the spectrum presented by Moran et al. (1996).  The emission-line flux
ratios places this galaxy on the boundary between regions ordinarily
populated by H{\sc II} and Seyfert galaxies. Although there are signs of
Seyfert activity as well (such as the presence of a weak, broad component
of the H$\alpha$ line), the optical spectrum is dominated by
characteristics of H{\sc II} galaxies. Moran et al. (1996) have suggested
that ``obscuration" effects may be responsible for the optical spectra of
``composite" galaxies. However, both the \xmm\ and \chandra\ observations
of \iras\ show no sign of any intrinsic absorption in this galaxy. There
is a possibility that the X-ray nucleus is totally obscured
(Compton-thick) in which case we are viewing the reflected component in
the backside of the torus.  The lack of variability between the \xmm and
the \chandra epochs (as well as the \rosat epoch)  could in principle
favor such an interpretation.  Nevertheless, the moderate equivalent width
of the Fe line, combined with the steep X-ray spectrum (see e.g. Matt et
al. 1996)  argue against the reflection component scenario. 
 The detection of a strong nuclear UV source and of a broad $H_{\alpha}$ 
 in the low signal-to-noise optical spectrum further support the
 hypothesis that we are viewing directly the Seyfert nucleus.
 Ward et al. (1988) have noted the presence of a tight  correlation
 between the broad $H\alpha$ and the hard X-ray (2-10 keV) 
 luminosity in unobscured AGN. Our object has an $H\alpha$ luminosity
 of $7\times 10^{40}$ \lunits (Moran et al. 1996) which translates to a 
 predicted 2-10 keV X-ray luminosity of $4\times 10^{42}$ \lunits, 
 in excellent agreement with the \chandra and \xmm observations, 
 further suggesting that the Broad Line Region is not obscured in our object.

One possibility that could explain the optical spectral properties of
\iras\ is that it hosts a non-standard broad line region. For some reason,
much less gas in the central source of this object is illuminated by the
ionizing continuum, than in a typical AGN. Or perhaps, the broad line
region may even be absent in this source. A third possibility though, and
perhaps a more realistic one, is that the AGN optical signature is not
evident in this source because of the presence of a starburst component
(Moran et al., 1996) {\it and} because it is partially diluted by the
presence of the strong optical continuum of the host galaxy. \iras\ is a
luminous spiral galaxy, as its optical/near-infrared excess emission in
the overall SED implies (Fig. 5). Its total K-band magnitude is
11.28 (2MASS survey) which translates to an absolute magnitude of
$M_{K}=-24.5$. Using an average colour of $\rm R-K=2.36$, appropriate for
spiral galaxies of morphological type $0\le T\le6$ (de Jong, 1996), we
find $M_{R}=-22.14$. Severgnini et al. (2003) estimate that the optical
signature of X-ray unabsorbed AGN with X-ray luminosity up to $\sim
10^{43}$ \lunits (the case of \iras), may be hidden if they are hosted by a
galaxy with $M_{R}\le -22$. In fact, their calculations indicate that only
H$\alpha$ will show clearly the appearence of the central active nucleus,
exactly like the case with \iras, as only H$\alpha$ shows clear evidence
for a broad component.

Finally, we do not detect any significant variations in both the \chandra\
and \xmm\ light curves. X-ray variability has become one of the defining
characteristics of AGN the last years. However, the lack of detection of
significant variations is not surprising, given the short duration of the
X-ray light curves. For example, the \xmm\ EPIC-PN light curve is less
than 10 ksec. For a $\sim 10^{43}$ erg/s source, we expect to measure an
``excess variance", $\sigma^{2}_{nxs}$ (i.e. the variance of a light curve
normalized by its mean squared after correcting for the experimental
noise)  of $\sim 0.01$ in light curves as long as $\sim 1$ day (e.g.
Nandra et al., 1997). If $\sigma^{2}_{nxs}$ increases proportional to the
length of the light curve, then we expect $\sigma^{2}_{nxs} \sim 0.0015$
in the case of the present \iras\ light curves. However, due to the low
count rate, even in the combined EPIC PN and MOS light curve, the
uncertainty on the estimation of $\sigma^{2}_{nxs}$ due to the
experimental Poisson noise (found using equation 11 of Vaughan et al.,
2003) is twice as large as the expected ``signal" that we want to detect.  
Clearly, a longer X-ray light curve is needed in order to detect
significant, short term variations in \iras. More surprising is the lack
of ``long term" variations, when we compare the observed X-ray flux during
the \xmm\, \chandra and \rosat observations, which are separated by 3
weeks and 12 years, respectively. Only monitoring X-ray observations over
a period of months/years could allow us to investigate the variability
behavior of this source.
 
\section{Summary} 

We present \chandra and \xmm observations of \iras\ one of the 'composite'
galaxies in the Moran et al. (1996) sample of galaxies.  The \chandra data
show that the X-ray emission comes from a nuclear point-like source with a
luminosity of $\rm L(2-10 keV) \sim 4\times 10^{42}$ \lunits. This
coincides with a bright, UV nucleus as revealed by the UV imaging
performed with the OM on-board \xmm. The X-ray spectrum (\xmm and
\chandra) is represented by a power-law spectrum with a steep photon index
$\Gamma\sim 1.7-1.9$ and a thermal component with 
 a temperature of $\sim 0.2$ keV absorbed by a negligible intrinsic column density
$<3\times 10^{20}$ \cunits. An Fe line at 6.4 keV is also detected in both the
\xmm and \chandra data with an equivalent width of $\sim 300$ eV. 
 The above give away the presence of an active nucleus. There is no
variability between the \chandra and the \xmm flux which are separated by a 
small period of about 3 weeks.  Puzzlingly, there is no observed
variability in the soft X-ray flux between \xmm and \chandra and the
\rosat observation which are separated by a period of more than 12 years.  
The marked lack of variability could suggest the presence of a highly
obscured (Compton-thick) AGN. However, this possibility is probably ruled
out by the moderate equivalent width of the Fe line, the steep X-ray
spectrum observed and the ratio of the hard X-ray to the $H\alpha$ 
luminosity which are typical of unobscured AGN.

The nuclear optical spectrum (using a $2''$ slit which corresponds to a
$\approx1.5$ kpc region at the source) obtained 2.5 months after the \xmm
observation, presents a weak broad $H\alpha$ line. Nevertheless, the
optical spectrum is dominated by a strong star-forming component which
outshines the AGN signatures. The weakness of the AGN features could be
attributed to the dilution of the nuclear AGN emission by the star-forming
component and the strong host galaxy optical emission.

\iras\ presents many similarities with the luminous X-ray galaxies
which have been detected in deep X-ray surveys. Although these are clearly
associated with AGN having high X-ray luminosities, their optical spectra
show no AGN signatures.  The present observations help us to shed light on
the properties of such objects and to understand why these fail to reveal
strong AGN signatures in the optical band.  This work demonstrates in
agreement with earlier work (Georgantopoulos et al. 2003, Severgnini et
al. 2003)  that heavy obscuration may not always be responsible for the
discrepancy observed between the X-ray and optical classification.

\acknowledgements This work has been supported by the NASA grant
G01-2120X. Support is also ackowledged by the grant 'X-ray Astrophysics
with ESA's mission XMM' in the framework of the programme 'Excellence in
Research' jointly funded by the the European Union and 
the Greek Ministry of Development. 
 Skinakas Observatory is a collaborative project of the University of Crete, the
Foundation for Research and Tecnology-Hellas, and the Max-Planck-Institut
f\"ur Extraterrestrische Physik. This research has made use of the
NASA/IPAC Extragalactic Database (NED) which is operated by the Jet
Propulsion Laboratory, California Institute of Technology, under contract
with the National Aeronautics and Space Administration.



\clearpage

\begin{table}
\caption{The spectral fits results}
\begin{tabular}{ccccccc}
\tableline\tableline
      & $\Gamma$ & $N_H$ & $E_1$  & $E_2$  & kT & $\chi^2$/dof \\
      &    & $10^{20}\rm cm^{-2}$ & keV & keV & keV &  \\
\tableline
PN &  $1.94^{+0.08}_{-0.10}$  & $4^{+3}_{-2}$ & 6.4 & 6.96 & $0.20^{+0.06}_{-0.02}$ & 233.3/186  \\
MOS&  $2.00^{+0.11}_{-0.09}$ &$5^{+2}_{-1}$ & -  & - & - & 128.5/121  \\
PN+MOS & $1.96^{+0.08}_{-0.06}$ & $3^{+2}_{-1}$ & 6.4 & 6.96 & $0.24^{+0.06}_{-0.06}$ & 347.9/307 \\     
ACIS-S & $1.67^{+0.05}_{-0.05}$ & $<2$ & 6.4 & - & $0.22^{+0.03}_{-0.03}$ & 246.3/199  \\

\tableline
\end{tabular}
\end{table}

\clearpage

\begin{inlinefigure}
\epsscale{1}
\rotatebox{0}{
\plotone{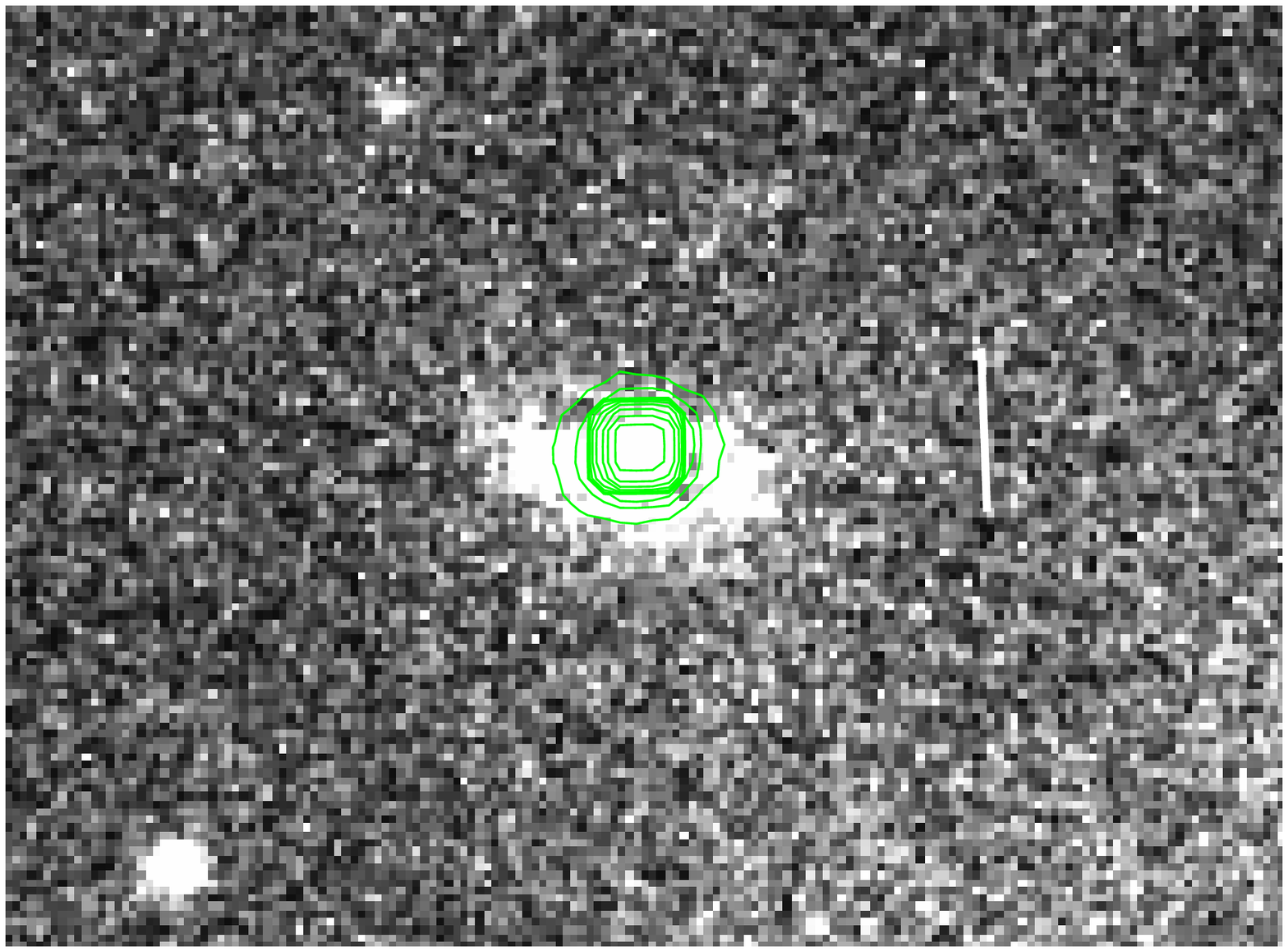}}
\figcaption{The X-ray contours from the 
 0.3-8 keV {\it Chandra} image,
  overlayed on the \xmm OM image (U filter).
 The line corresponds to $10''$.}
\end{inlinefigure}

\begin{inlinefigure}
\epsscale{0.6}
\rotatebox{0}{
\plotone{f2.eps}}
\figcaption{Upper panel: \xmm EPIC-PN, MOS-1 and MOS-2, background subtracted,
$0.3-8$ keV light curves of \iras, grouped in 200 s bins (open spqaures,
filled and open circles respectively). Time is counted from the start of
the PN observation (for clarity reasons the points have been sifted by
+50 s on the time axis). Lower panel: \chandra, background subtracted,
$0.3-8$ keV light curves of \iras, grouped in 1500 s bins. Time is
counted from the start of the \chandra observation (for clarity reasons
the points have been sifted by +300 s sec on the time axis).}
\end{inlinefigure}

\begin{inlinefigure}
\epsscale{0.6}
\rotatebox{270}{
\plotone{f3a.eps}}
\rotatebox{270}{\plotone{f3b.eps}}
\figcaption{The \xmm PN (upper panel)
 and \chandra (lower panel) spectra    
 plotted together with a 
 model fit consisting of a power-law plus 
a 6.4 keV Gaussian line, absorbed 
by the Galactic and an intrinsic column density. 
The data to model ratios are given as well.}
\end{inlinefigure}

\begin{inlinefigure}
\epsscale{0.8}
\rotatebox{0}{
\plotone{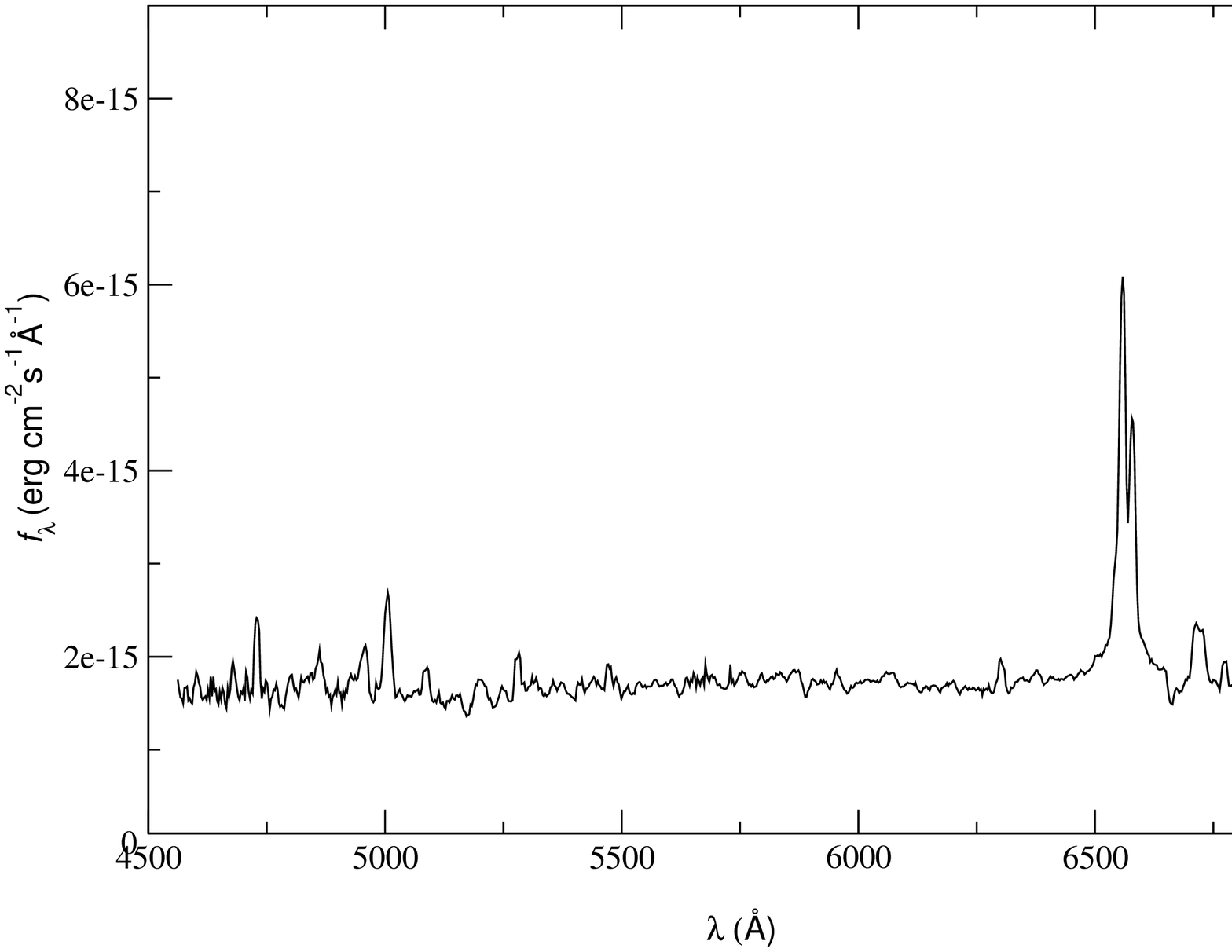}}
\figcaption{The optical spectrum of \iras\ taken in June 2002, from
Skinakas observatory. The spectrum, plotted as flux per unit wavelength
interval, has been shifted to rest frame.}
\end{inlinefigure}

\begin{inlinefigure}
\epsscale{0.8}
\rotatebox{0}{
\plotone{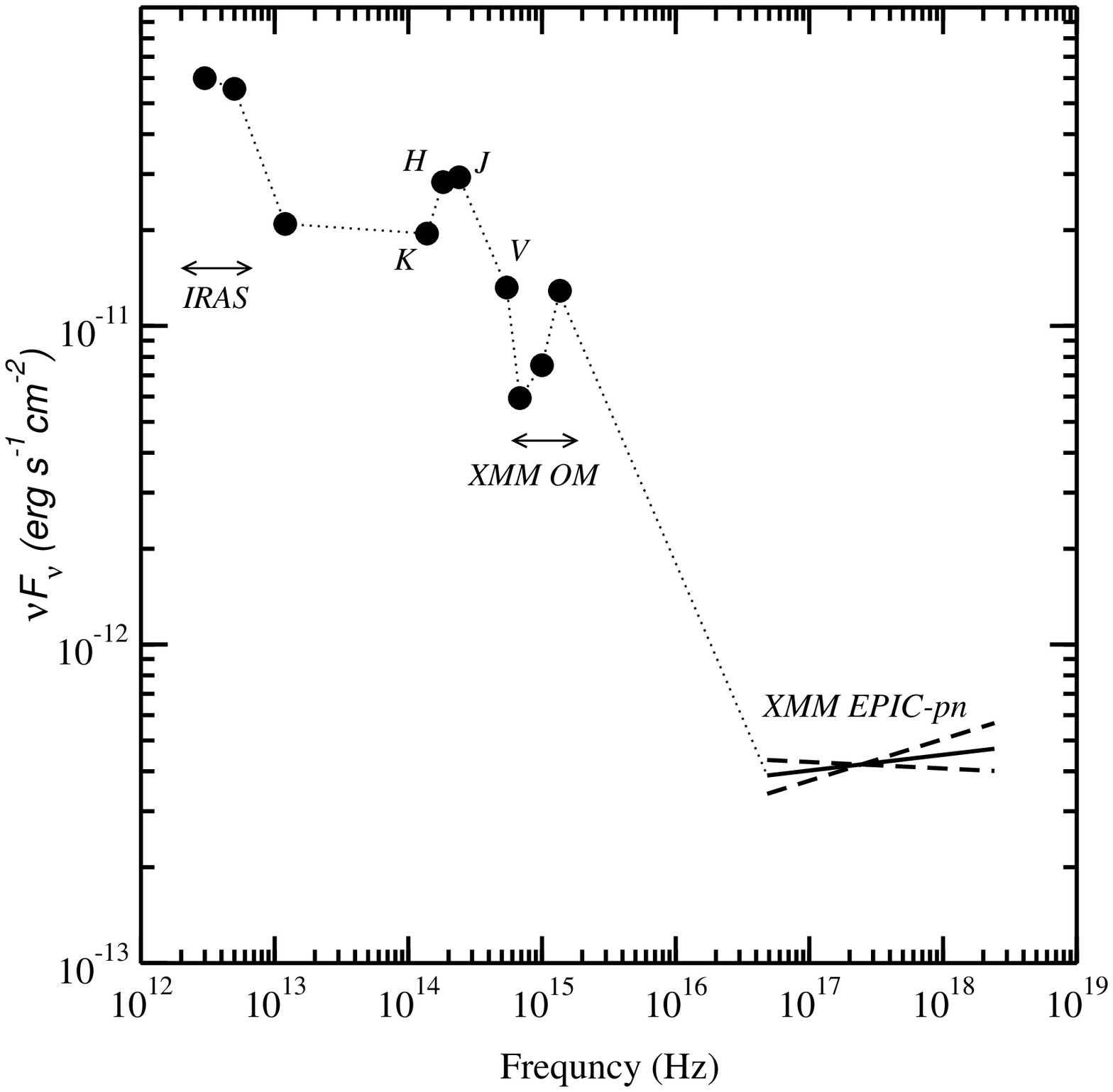}}
\figcaption{The rest frame, dereddened continuum energy distribution of 
\iras. Note that the apertures used to derive fluxes in the various bands do
not have the same size (see text for details).}
\end{inlinefigure}


\begin{references}

\reference{} Alexander, D.M. et al.  2003, AJ, 126, 539
\reference{} Barger, A.J., Cowie, L.L., Mushotzky, R.F., Richards,
E.A., 2001, AJ, 121, 662
\reference{} Bohlin, R.C., Savage, B. D., Drake, J.F., 1978, \apj, 224, 132 
\reference{} Boyle, B.J., McMahon, R.G., Wilkes, B.J., Elvis, M.,
 1995, MNRAS, 276, 315 
\reference{} Cardelli, J.A., Clayton, G.C., Mathis, J.S., 1989, \apj, 345, 245
\reference{} Comastri, A., et al., \apj, 571, 771   
\reference{} de Jong, R.S., 1996, A\&A, 313, 377 
\reference{} Dickey, J.M., Lockman, F.J., 1990, ARA\&A, 28, 215
\reference{} Elvis, M., et al., 1994, \apjs, 95, 1 
\reference{} Georgantopoulos, I., Zezas, A., Ward, M.J., 2003, \apj, 584, 129
\reference{} Green, P.J. et al. 2004, ApJS, 150, 43 
\reference{} Griffiths, R.E., Georgantopoulos, I., Boyle, B. J., 
 Stewart, G.C., Shanks, T., della Ceca, R., 1995, MNRAS, 275, 77
\reference{} Matt, G., et al. 1996, 281, L69  
\reference{} McHardy, I.M., et al., 1998, MNRAS, 295, 641 
\reference{} Misiriotis A., Papadakis I.E., Kylafis N.D., Papamastorakis J., 2004,
A\&A, 417, 39 
\reference{} Moran, E.C., Halpern, J.P., Helfand, D.J., 1996, \apjs, 106, 341
\reference{} Moran, E.C., Lehnert, M.D., Helfand, D.J., 1999, \apj, 526, 649 
\reference{} Moran, E.C., Filippenko, A.V., Chornock, R., 2002, \apj, 579, L71
\reference{} Mushotzky, R.F., Cowie, L.L., Barger, A.J., Arnaud, K. A. 
 2000, Nature, 404, 459
\reference{} Nandra, K., George, I.M., Mushotzky, R.F., Turner, T.J., 
 Yaqoob, T. 1997, \apj, 476, 70 
\reference{} Page, K. L., O'Brien, P.T., Reeves, J. N., Turner, M.J.L., 
 2004, MNRAS, 347, 316  
\reference{} Schlegel, D.J., Finkbeiner, D.P., Davis, M., 1998, \apj, 500, 525 
\reference{} Severgnini, P. et al. 2003, A\&A, 406, 483 
\reference{} Stevens, I.R., Read, A.M., Bravo-Guerrero, J., 2003, MNRAS,
 343, L47
\reference{} Stocke, J., et al. 1991, \apjs, 76, 813 
\reference{} Vaughan, S., Edelson, R., Warwick, R.S., Uttley, P., 2003, MNRAS, 345, 1271 
\reference{} Veilleux S., Kim D.-C., Sanders D.B., Mazzarella J.M.,
Soifer B.T.,  1995, \apjs, 98, 171 
\reference{} Veron, P., Veron, M.P., Bergeron, J., Zuidervijk, E.J.,
 1981, A\&A, 97, 71 
\reference{} Ward, M. J., Done, C., Fabian, A.C., Tennant, A.F., Shafer, R.A., 1988, \apj, 324, 767
\reference{} Watson, M.G., 2001, A\&A, 365, L51
\reference{} Weisskopf, M.C., O'dell, S.L., van Speybroeck, Leon P.,
1996, SPIE, 2805, 2  
\reference{} Zezas, A., Georgantopoulos, I., Ward, M.J., 1998, MNRAS, 301, 915 


\end{references}
\end{document}